\documentclass[11pt]{article}
\def\la#1{\hbox to #1pc{\leftarrowfill}}
\def\ra#1{\hbox to #1pc{\rightarrowfill}}
\def\fract#1#2{\raise4pt\hbox{$ #1 \atop #2 $}}
\def\decdnar#1{\phantom{\hbox{$\scriptstyle{#1}$}}
\left\downarrow\vbox{\vskip15pt\hbox{$\scriptstyle{#1}$}}\right.}

\def\la#1{\hbox to #1pc{\leftarrowfill}}
\def\ra#1{\hbox to #1pc{\rightarrowfill}}
\def\fract#1#2{\raise4pt\hbox{$ #1 \atop #2 $}}
\def\decdnar#1{\phantom{\hbox{$\scriptstyle{#1}$}}
\left\downarrow\vbox{\vskip10pt\hbox{$\scriptstyle{#1}$}}\right.}

\font\sc=cmcsc10 at 12pt

\def\za{\vrule height6pt width4pt depth1pt}
\def\lrar{{\ra 2}}

\def\map#1{\hbox{Map}_{#1}}
\def\bmap#1{\hbox{Map}^*_{#1}}
\def\rat#1{\hbox{Rat}_{#1}}
\def\hol#1{\hbox{Hol}_{#1}}
\def\bhol#1{\hbox{Hol}^*_{#1}}
\def\d#1{\hbox{Div}_{#1}}

\def\sp#1{SP^{#1}}
\def\spy{SP^{\infty}}
\def\map#1{\hbox{Map}_{#1}}

\def\tensor{\otimes}
\def\deg{\hbox{deg}}
\def\mod{\hbox{Mod}}

\def\bbz{{\bf Z}}
\def\bbf{{\bf F}}
\def\bbp{{\bf P}}
\def\bbr{{\bf R}}
\def\bba{{\bf A}}
\def\bbc{{\bf C}}
\def\bbr{{\bf R}}

\title{\huge Configuration Spaces and the\\
Topology of Curves in Projective Space}

\author{Sadok Kallel\\
{\small\it Laboratoire AGAT,
Universit\'e Lille I, France
\thanks{sadok.kallel@agat.univ-lille1.fr}}}
\date{}

\begin{document}
\maketitle


\begin{abstract}
We survey and expand on the work of Segal, Milgram and the author on the
topology of spaces of maps of positive genus curves into $n$-th
complex projective space, $n\geq 1$ (in both the holomorphic and
continuous categories).  Both based and unbased maps are studied and
in particular we compute the fundamental groups of the spaces in
question. The relevant case when $n=1$ is given by a non-trivial
extension which we fully determine.
\end{abstract}


\noindent{\bf\large\S1 Introduction and Statement of Results}
\vskip 10pt

The topology of spaces of rational maps into various complex manifolds
has been extensively studied in the past two decades. Initiated by
work of Segal and Brockett in control theory, and then motivated by
work of Donaldson in gauge theory, this study has uncovered some
beautiful phenomena (cf.  [CM], [Hu], [L]) and brought to light
interesting relationships between various areas of mathematics and
physics (cf. [Hi], [BM],[BHMM]). 

Let $C_g$ (or simply $C$) denote a genus $g$ (compact) Riemann surface
($\bbp^1$ when $g=0$) and let $V$ be a complex projective
variety. Both $C$ and $V$ come with the choice of preferred basepoints
$x_0$ and $*$ respectively.  The main focus of this paper is the study
of the geometry of the space
$$\hol{}(C, V)=\{f: C\lrar V, f~\hbox{holomorphic}\}$$
and more particularly its subspace of basepoint preserving maps
$$\hol{}^*(C, V)=\{f: C\lrar V, f~\hbox{holomorphic and}~ f(x_0) =
*\}$$ The space $\hol{}^*(C, V)$ doesn't depend up to homeomorphism
on the choice of basepoint when $V$ has a transitive group of
automorphisms for example. In that situation, the relationship between
the based and unbased (or free) mapping spaces is given by the
``evaluation''
$$\hol{}^*(C, V)\lrar\hol{}(C, V)\fract{ev}{\lrar} V$$
which is a holomorphic fibration for $V$ a homogeneous 
space for example. Here $ev$ evaluates a map at the basepoint of $C$. 

\noindent{\bf Remark} 1.1: 
By choosing a Riemann surface, we fix the complex structure.  We
record the link with the theory of ``pseudo-holomorphic'' curves of
Gromov. There one studies the (differential) geometry of spaces of
$J$-holomorphic curves into $V$, where $V$ is almost complex and $J$
is its almost complex structure. These $J$-holomorphic curves (of
given genus $g$) correspond when $J$ is integrable to all holomorphic
maps of $C$ (with varying complex structure) into $V$. It is now a
theorem of Gromov that for a generic choice of $J$ on $V$, the space
of all such curves (in a given homology class) is a smooth finite
dimensional (real) manifold. Let ${\cal M}(V,g)=\{(C_g, f)~|~f: C\lrar
V\ \hbox{holomorphic}\}$, then there is a projection of ${\cal M}(V,g)$,
onto the moduli space of curves ${\cal M}_g$ sending $(C, u)$ to the
isomorphism class of $C$. The spaces $\hol{}(C, V)$ appear then as
``fibers'' of this projection.

We observe that $\hol{}(C, V)$ breaks down into connected components
and we write $\hol{A}(C, V)$ for the component of maps $f$ such that
$f_*[C]=A$, $A$ a given homology class in $H_2(V)$. When $V$ is simply
connected, $[C,V]=[S^2,V]=\pi_2(V)=H_2(V)={\bf Z}^r$ for some rank
$r$ and $A$ is completely determined by a multi-degree.  The
components $\hol{A}(C, V)$ are generally (singular) quasi-projective
varieties (see [H2] for example). When $V=\bbp^n$, we write $\hol{A}(C,
\bbp^n) = \hol{k}(C, \bbp^n)$ for some $k=\deg A\in \bbz$ and these
are smooth manifolds as soon as $k$ is big enough (about twice the
genus of $C$). Most of this paper is concerned with the study of these
spaces.

\noindent{\bf Remark} 1.2: In physics, spaces of maps between
manifolds $\hol{}(M,V)\subset \map{}(M,V)$ arise in connection with
field theory or ``sigma models''. 
From the perspective of a physicist, a field on
$M$ with values in $V$ is a map $\phi: M\rightarrow V$. For example,
in the case $M={\bf R}^3$ and $V={\bf S}^3=\bbr^3\cup\{\infty\}$, a
map $\phi:{\bf R}^3\rightarrow S^3$ could be the field associated to
some electrical charge in ${\bf R}^3$ (hence vector valued and
extending to the point where the charge is located by mapping to
$\infty$).  Associated to a field there is an ``energy'' density or
Lagrangian $\cal L$ (eg. the harmonic measure ${\cal L}(\phi )={1\over
2}||d(\phi )||^2$).  To an energy density one can in turn associate an
``action'' which is defined as
$$S[\phi ] = \int_{M}{\cal L}(\phi )d\mu (h)$$ 
where $d\mu (h)$ is the canonical volume measure associated to a
metric $h$ on $M$. Physicists are usually interested in minimizing the
action (to determine the dynamics of the system)
and hence they are led to study the space of all extrema of this
functional. It should be noted that in the case when $V$ is compact,
Kahler, a well-known theorem of Eells and Wood asserts that the
absolute minima of the energy functional on $\map{}(C,V)$ are the
holomorphic maps (the critical points here being the harmonic maps).

It has been known since the work of Segal [S] that the topology of
holomorphic maps of a given degree $k\in{\bf N}$ from $C$ to
$\bbp^n$ compares well with the space of continous maps at least
through a range increasing with $k$. He proved

\noindent{\bf Theorem} 1.3: (Segal)~{\sl The inclusion
$$\bhol{k}(C,\bbp^n)\hookrightarrow\bmap{k}(C,\bbp^n )$$
induces homology isomorphisms up to dimension $(k-2g)(2n-1)$}

A similar statement holds for unbased maps. To simplify notation, we
write $\map{k}$, $\hol{k}$, $\bmap{k}$, $\bhol{k}$ for the corresponding
mapping spaces from $C$ into $\bbp^n$ .

When $g=0$ (the rational case) and $n>1$ the homology isomorphism in
1.3 can be upgraded to a homotopy equivalence (cf. [CS]).  For $g>0$
however it is not known whether the equivalence in 1.3 holds in
homotopy as well; i.e whether the pair $(\map{k}, \hol{k})$ is
actually $(k-2g)(2n-1)$ connected. This is strongly suspected to be
true and in this note we give further evidence for this by showing
that $\bhol{k}$ and $\bmap{k}$ have isomorphic fundamental groups for
all $n$ (the relevant case here is $n=1$). In fact we shall show

\noindent{\bf Theorem} 1.4:~{\sl Suppose $k\geq 2g$, we have
isomorphisms
$$\pi_1(\bhol{k}(C_g,\bbp^n ))\cong \pi_1(\bmap{0}(C_g,\bbp^n ))
\cong \left\{{\bbz^{2g},\ \hbox{when}\  n>1}\atop
{G,\ \hbox{when}\  n=1}\right.$$
where $G$ is a cyclic extension of $\bbz^{2g}$ by $\bbz$
generated by classes 
$e_1,\ldots, e_{2g}$ and $\tau$ such that the commutators
$$[e_i, e_{g+i}] = \tau^2$$
and all other commutators are zero. }

\noindent{\sc Remark}\footnote{J.D.S.Jones has recently informed the 
author that he has obtained a similar result a few years ago but which
he didn't publish.}: Roughly speaking, the class $\alpha$ can be
represented by the one parameter family of maps obtained by rotating
roots around poles. Similarly the classes $e_i$ are obtained by
rotating roots (or poles) around loops representing the homology
generators of $C$.

Combining this result with a classical result of G. Whitehead (see
\S2), we deduce

\noindent{\bf Proposition} 1.5:~{\sl
$\pi_1(\map{d}(C,S^2 ))$ is generated by classes 
$e_1,\ldots, e_{2g}$ and $\alpha$ such that 
$$\alpha^{2|d|}=1,~~ [e_i, e_{g+i}] = \alpha^2$$
and all other commutators are zero. When $d\geq 2g$,
$\pi_1(\hol{d}(C, \bbp^1))\cong \pi_1(\map{d}(C,S^2 )).$}

This description for the continuous mapping space is an earlier result
of Larmore-Thomas [LT].  As is clear from 1.5, the components of
$\map{}(C,\bbp^n )$ for $n=1$ have different homotopy types. When
$n>1$, the fundamental group is however not enough to distinguish
between the components. A quick byproduct of our calculations however
shows (\S8)

\noindent{\bf Proposition} 1.6:~{\sl $\map{k}(C,\bbp^{2d})$ and
$\map{l}(C,\bbp^{2d})$ have different homotopy types whenever
$l$ and $k$ have different parity\footnote{it is now verified that
all (positive) components of $\map{}(C,\bbp^n)$ have different
homotopy type; cf. [KS].}.}

The first few sections of this paper are written in a leisurely
fashion and are meant in part to survey techniques and ideas in the
field most of which carry the deep imprint of Jim Milgram (cf. \S4).  

In \S2 we discuss the rational case (i.e. $g=0$) and give a short
proof of a theorem of Havlicek on the homology of unbased
self-morphisms of the sphere.  In \S3 we introduce a configuration
space model (originally given in [K2]) for $\bmap{}(C,\bbp^n )$ and
use it to determine the homology of this mapping space.  Let
$\sp{k}(M)= M^k/\Sigma_k$ where $\Sigma_k$ is the cyclic group on
$k$-letters acting by permutations (or equivalently the space of
unordered $k$ points on $M$), and let $\sp{k}_n(M)$ be the subspace of
$k$ points in $C$ no more than $n$ of which are the same ($n\leq
k$). When $n=1$, $\sp{k}_1(M)=C_k(M)$ is the standard configuration
space of $k$ distinct points. When $M$ is a (connected) open manifold
or with a (collared) boundary, these spaces can be stabilized
(cf. \S3) $\sp{k}_n(M)\lrar\sp{k+1}_n(M)\lrar\cdots $ and we write
$\spy_n(M)$ for the (connected) direct limit.  The following is
discussed in \S3

\noindent{\bf Theorem} 1.7 [K2]: {\sl There is a map
$$\spy_n(C-*)\fract{S}{\ra 3}\bmap{0}(C,\bbp^{n})$$
which is a homotopy equivalence when $n>1$ and a homology
equivalence when $n=1$.}

This model for $\bmap{}$ in terms of `configurations of bounded
multiplicity' turns out to relate quite well with the corresponding
model for $\bhol{}$ and we indicate based on this another proof
for Segal's result (\S5). Notice that 1.4 shows that the homology
equivalence $S$ above cannot be upgraded to a homotopy equivalence
when $n=1$ for already both spaces have different fundamental
groups. Indeed the braid group $\pi_1(\spy_1(C_g-*))=\pi_1(C(C_g-*))$
has a presentation quite different from the central extension given in
1.4.

\vskip 10pt
\noindent{\sc Acknowledgements}: We would like to take this occasion
to deeply thank Jim Milgram for the many beautiful mathematics we have
learned from him. We thank J. F. Barraud, Y. Hantout and J. Nagel for
useful discussions we had. Special thanks to the PIms institute in
Vancouver for its support while part of this work was carried out.

\vskip 10pt
\noindent{\bf\Large\S2 Preliminaries: The Genus Zero Case}
\vskip 10pt

It is customary to write $\rat{}(V)$ for the basepoint preserving maps
$\hol{}^*(S^2,V)$. When $V=\bbp^n$, these spaces are now very well
understood. Let
$$C_k(\bbc, S^{2n-1})=\coprod_{0\leq i\leq
k}F(\bbc,i)\times_{\Sigma_i}(S^{2n-1})^i/\sim$$ be the standard
labeled $k$-th filtration piece for the May-Milgram model
of $\Omega^2S^{2n+1}\simeq
\Omega^2_0(\bbp^n)$ (cf. [C$^2$M$^2$], [K4], etc). Here $F(\bbc, i)$
is the set of ordered $i$-tuples of disctinct points in $\bbc$ and
$\sim$ is a standard basepoint identification identifying an $i$-tuple
with labels to an $i-1$ tuple by discarding a point an its label 
if the label is at basepoint. Let
$$\sp{k}(X)=\{\sum k_ix_i, x_i\in X, k_i\in {\bf N}\ |\ 
x_i\neq x_j~\hbox{for}~ i\neq j~\hbox{and}~\sum k_i=k\}$$ 
be the $k$-th symmetric product of $X$ (see introduction). Let 
$\sp{k}_n(X)$ be the subset of $\sp{k}(X)$ obtained by restricting
to $k_i\leq n$. We then have

\noindent{\bf Theorem} 2.1 ([CS], [K3], [GKY]):~{\sl For $k\geq
1$, there are maps and homotopy equivalences
$$\rat{k}(\bbp^n)\fract{\simeq}{\la 3}
C_k(\bbc, S^{2n-1})\fract{\simeq}{\ra 3}\sp{k(n+1)}_n(\bbc )
\leqno{2.1(a)}$$
whenever $n>1$. When $n=1$ the spaces are homologous only}.

It is interesting to note that no map is known to induce such homology
isomorphism between $\rat{k}(\bbp^n )$ and $\sp{k(n+1)}_n(\bbc )$ ($k>1$).
When $n>1$, the left-hand map in 2.1(a) is constructed explicitly in
[CS] while the right-hand map is constructed in [K3].  Both
configuration space models on either side of 2.1(a) are fairly
amenable to calculations and from there the structure of
$\rat{k}(\bbp^n)$ can be made quite explicit (cf. [BM], [C$^2$M$^2$],
[K3] and [Ka]).

The space of unbased rational maps is less well understood. The following
is due to Havlicek ([H1])

\noindent{\bf Theorem} 2.2:~{\sl
The Serre spectral sequence for the (holomorphic) fiber bundle
$$\rat{k}({\bf P}^1)\lrar\hol{k}({\bf P}^1)\lrar{\bf P}^1\leqno{2.2(a)}$$
has the non-zero differential $d_2(x) = 2k\iota$.  
The spectral sequence collapses with mod-$p$
coefficients whenever $p=2$ or $p$ divides $k$.}

We give a short proof for the mod-2 collapse (a general proof and an
extension of this result to $\rat{k}(\bbp^n )$ can be found in
[KS]). First of all, to see that 2.2(a) is indeed a fiber bundle, one
simply observes that PSL$_2(\bbc )$ acts transitively on $\bbp^1$ and
if $F$ denotes the stabilizer of a point, then $F$ acts on
$\rat{k}({\bf P}^1)$ (by postcomposition). One can see that
$\hol{k}({\bf P}^1) = \rat{k}({\bf P}^1)\times_F\bbp^1$.

Towards the proof of 2.2 (and also in most of \S7) we need the following
classical result of G. Whitehead. Let $X$ be a based (connected)
topological space (with basepoint $x_0$) and consider the evaluation
fibration
$$\Omega^n_fX\lrar {\cal L}^n_f X\fract{ev}{\lrar} X\leqno{2.3}$$
where $ev(f)=f(x_0)$, ${\cal L}^n_fX=\map{f}(S^n,X)$ is the
component of the total mapping space containing a given map
$f:S^n\rightarrow X$ and $\Omega^n_fX$ the subset of all maps $g$ such
that $g(x_0)=f(x_0)$.

\noindent{\bf Theorem} 2.4 [W]:~{\sl The homotopy boundary in 2.3
$\partial:\pi_i(X)\rightarrow\pi_i(\Omega^n_f X)\cong\pi_{i+n}(X)$ is
given (up to sign) by the Whitehead product: $\partial \alpha =
[\alpha,f].$}

Let ${\cal L}^2_kS^2={\cal L}_f^2S^2$ where $f$ is the standard degree $k$
map, and denote by $a\in H^2(S^2)$ the generator. Recall that
$\Omega^2S^2\simeq \bbz\times\Omega^2S^3$ and so let $e$ be the generator in
$H^1(\Omega^2_kS^2)\cong H^1(\Omega^2S^3)$.

\noindent{\bf Lemma} 2.5:~{\sl The Serre spectral sequence for
$\Omega^2_kS^2\lrar{\cal L}^2_kS^2\lrar S^2$ has the (homology)
differential $d_2(a) = 2ke$. It collapses at $E_2$ with mod-2
coefficients.}

\noindent{\sc Proof}: For a fibration
$F\rightarrow E\fract{p}{\lrar} S^2$ we have the diagram
$$\begin{array}{cccc}
  {\pi_2(E,F)\atop\cong\pi_2(S^2)}&
  \fract{\partial}{\lrar}&\pi_{1}(F)\\
  \decdnar{h}&&\decdnar{h}\\
  H_2(E,F)&
  \fract{\partial}{\lrar}&H_{1}(F)\\
  \decdnar{p_*}&&\decdnar{}\\
  H_2(S^2)&\fract{\tau}{\lrar}&E^{0,1}
\end{array}
$$
where $E^{0,1}\cong H_{1}(F)$ and $\tau$ the transgression.
Let $k:S^2\lrar S^2$ denote multiplication by $k$. From 2.4 and
the diagram above,
we deduce that
$$\tau: H_2(S^2)\cong \pi_2(S^2)\fract{k[\iota,- ]}{\ra 3}\pi_3(S^2)
\fract{ad}{\lrar}\pi_1(\Omega^2_kS^2)\fract{h}{\lrar}
H_1(\Omega^2_kS^2)=\bbz$$
$\iota\in \pi_2(S^2)$ is the generator.
As is known for even spheres, $h(ad ([\iota,\iota ]))=2$
and this yields the differential.

The mod-2 collapse follows from the following very short argument
(which we attribute to Fred Cohen): let $E:S^2\lrar\Omega S^3$ be the
adjoint map. Then there is a map of (horizontal) fibrations
$$\begin{array}{cccccc}
\Omega^2_kS^2&\lrar&{\cal L}^2_kS^2&\fract{ev}{\lrar}& S^2\\
\decdnar{\Omega^2E}&&\decdnar{}&&\decdnar{E}\\
\Omega^3_kS^3&\lrar&{\cal L}_k^2\Omega S^3&\lrar&\Omega S^3
\end{array}
$$
The map $E$ is injective in integral homology while $\Omega^2E$ is
injective in mod-2 homology. So mod-2, both fiber and base inject in
the bottom fibration which is trivial since $\Omega S^3$ is (homotopy
equivalent) to a topological group (and translation of basepoint
gives the trivialization). 
The collapse follows in this case.  \hfill\za

\vskip 10pt
\noindent{\sc Proof of 2.2}: We have inclusions
$\rat{}(\bbp^1)\subset \hol{}(\bbp^1)\subset \Omega^2S^2$, where
$\rat{}(\bbp^1)$ is the subspace of all holomorphic maps sending the
north pole $\infty$ in $\bbp^1=\bbc\cup{\infty}$ to 1.  An element of
$\rat{}(\bbp^1)$, say of degree $k$, is given as a quotient ${p\over
q}={{z^k+a_{k-1}z^{k-1}+ \cdots + a_0}\over {z^k+b_{k-1}z^{k-1}+\cdots
+ b_0}}$ where $p$ and $q$ have no roots in common. It is easy to see
for example that $\hol{1}(\bbp^1)$ corresponds to $PSL(2,\bbc )$, the
automorphism group of $\bbp^1$ (which is up to homotopy $RP^3$), and
that $\rat{1}(\bbp^1)=\bbc\times\bbc^*\simeq S^1$.
Consider now the map of fibrations
$$\begin{array}{ccc}
\rat{k}(\bbp^1)&\hookrightarrow&\Omega^2_kS^2\\
\downarrow&&\downarrow\\
\hol{k}(\bbp^1)&\hookrightarrow&{\cal L}^2_kS^2\\
\downarrow&&\downarrow\\
{\bf P}^1&\fract{=}{\lrar}&S^2
\end{array}
\leqno{2.6}$$
According to Segal (theorem 1.3), the top inclusion is an isomorphism
in homology up to dimension $k$. Moreover (cf. [C$^2$M$^2$], [K1]),
$H_*(\rat{k}(\bbp^1))$ actually injects in $H_*(\Omega^2_0S^2)$ and it
does so in the following nice way.  Recall that
$\Omega^2_0S^2\simeq\Omega^2S^3=\Omega^2\Sigma^2S^1$ and hence it
stably splits as an infinite wedge $\bigvee_{j\geq 1}D_j$ where the
summands $D_j$ are given in terms of configuration spaces with
labels. Stably one finds that
$$\displaystyle \rat{k}(\bbp^1)\simeq_s\bigvee_{j=1}^kD_j.$$
The map of fibers in 2.6 is then an injection and since
the bases are same in that diagram, the theorem follows from 2.5 and a
spectral sequence comparison argument.\hfill\za

\vskip 10pt
\noindent{\bf\Large\S3 Configuration Space Models and the Higher Genus Case}
\vskip 10pt

In this section we describe two configuration space models for each of
the mapping spaces $\hol{k}^*(C,\bbp^n )$ and $\map{k}^*(C,\bbp^n )$
for $g\geq 1$. A straightforward comparison between both models yields
Segal's stability result in \S4.

First of all, a map $f\in \hol{k}(C,\bbp^n )$ can be written
locally in the form
$$z\mapsto [p_0(z):\ldots :p_n(z)]$$
where the $p_i(z)$ are polynomials of degree $k$ each and having
no roots in common. The $p_i$'s are not global functions on $C$
(otherwise they would be constant)  but rather local maps into $\bbc$
and hence sections of some line bundle.  The roots
of each $p_i$ give rise to an element $D_i\in\sp{k}(C)$ (so called
{\sl positive divisor}) and conversely $D_i$ only
determines $p_i$ up to a non-zero constant (which will be determined
if the maps are based).
These $D_i$'s ({\sl the root data}) cannot of course have
a root in common and if we base our maps so that basepoint $*$ is sent
to $[1:\cdots :1]$ say, then none of the $D_i$'s contains the
basepoint.  Define
$$\d{k}^{n+1}(X) =\{(D_0,\ldots, D_n)\in
\sp{k}(X)^{n+1}~|~D_0\cap\cdots\cap D_n=\emptyset\},$$
The previous discussion shows the existence of an inclusion
$$\phi: \hol{k}^*(C,\bbp^n)\lrar\d{k}^{n+1}(C-*)$$ 
which associates to a holomorphic map its root data.  This
correspondence has no inverse since it is not true in general that an
(n+1) tuple of degree $k$-divisors with no roots in common gives rise
to a (based) holomorphic map out of $C $. In fact, this can be
understood as follows: the quotients ${p_i(z)\over p_j(z)}$ are
meromorphic maps on $C$ and so for $i\neq j$, $D_i-D_j$ must be the
divisor of a function on $C$ (i.e. there is $f:C\rightarrow\bbc$ with
roots at $D_i$ and poles at $D_j$). It is not surprising that this
last condition is not satisfied for general pairs $(D_i,D_j)$.  We say
$D_i$ and $D_j$ are {\sl linearly equivalent} if there is indeed a
meromorphic function $f$ on $C$ such that $(f) := $zeroes of $f$-poles
of $f=D_i-D_j$. Note that when this is the case $D_i$ and $D_j$ have
same degree.

Linear equivalence defines an equivalence relation and one denotes by
$J(C)$ the set of all linearly equivalent divisors on $C$ of degree
$0$. There is a map
$$\mu:\sp{n}(C)\lrar J(C),~~ D\mapsto [D-nx_0]\leqno{3.1}$$
sending a divisor to the corresponding equivalence class.
The above discussion then shows the existence of a homeomorphism
$$
\left\{{\hbox{A (based) holomorphic}\atop
\hbox{$X\lrar\bbp^r$ of degree $k$}}\right\}
\Longleftrightarrow
\left\{{{{\hbox{An $(n+1)$ tuple of positive divisors }\atop
\hbox{$D_i$ on $C-*~|~D_0\cap\cdots\cap D_n=\emptyset$}}
\atop
\hbox{$\deg D_i=k$ and $\mu (D_i)=\mu (D_j)$ }}}\right\}$$
To make this correspondence more precise, we need understand the map
$\mu$ in 3.1. It turns out that:\hfill\break
$\bullet$ 
$J(C)$ is a $g$ dimensional complex torus (this is a non-trivial fact)
and $\mu$ is a multiplicative map.  In the case $g=1$, $\mu$ is
the identity and if one identifies the
curve with $\bbc/L$ for some lattice $L$, then the Abel-Jacobi condition
$\mu (\sum z_i )=\mu (\sum p_j ), z_i,p_i\in C$ is equivalent to
$\sum z_i = \sum p_j\mod (L)$ ($z$ for zero and $p$ for pole). 
\hfill\break
$\bullet$ 
The preimage of a point $[D]=\mu (D)\in J$ is a projective space. To
see this let ${\cal L}(D)$ be the set of all holomorphic maps on $C$
such that $(f)+D\geq 0$ (i.e. such that
$(f)+D\in\bigsqcup_n\sp{n}(C)$).  This is a $\bbc$-vector space and
since $(\alpha f)=(f)$, we have an identification and a map
$\mu^{-1}([D])=\bbp{\cal L}(D)\lrar\sp{n}(C),~[f]\mapsto (f)+ D$.
This turns out to be a holomorphic embedding ([G],\S4,
4.3). \hfill\break
$\bullet$
The complex dimension of $\mu^{-1}([D])$ is denoted by $r(D)$
and a crucial aspect of the classical theory of algebraic curves is
the computation of $r(D)$ as $D$ varies in $\sp{n}(C)$. A very
useful interpretation of this dimension is as follows: {\it
$r(D)\geq r$ if and only for any $r$ points of $C$, there is 
a divisor $D'$ with $\mu (D') =\mu (D)$ and passing through 
these $r$ points.}  \hfill\break
$\bullet$
If $\deg D=n$, then generically
$$r(D) = max\{0, n-g\}.$$
The exact dimension for every $D$ is determined by Riemann-Roch.  This
dimension may jump up for certain ``special" divisors. However
whenever $n$ exceeds $2g-2$, there are no more jumps and $r$ is
uniformally given by $n-g$.

It turns out that for $k$ small (i.e. $k< g$), $\hol{k}^*(C,{\bf P}^n)$ is
very dependent on the holomorphic structure put on the curve (it can
naturally be empty). For example if $C$ is hyperelliptic, then
$$\hol{2n+1}(C,\bbp^1)=\emptyset, 2n+1\leq g~~\hbox{(see for instance
[FK], Chap3)}$$
Also $\hol{1}(C,\bbp^n )$ is always empty for all positive genus curves.
In the range $k\geq 2g-1$, there is however much better behavior (see [KM])

\noindent{\bf Lemma} 3.2:~{\sl Assume $k\geq 2g-1$. Then $\hol{k}^*(C,
{\bf P}^n)$ is a $k(n+1)-ng$ complex manifold.}

\noindent{\sc Remark} 3.3: $\hol{k}^*(C, {\bf P}^n)$ has additionally the
structure of a quasiprojective variety.  Generally, components of
$\hol{}(C, V)$ do not have a smooth structure (when $V$ is a smooth
projective variety). This already fails for $V =G(n,n+k)$ (the
Grassmaniann of $n$ planes in $\bbc^{n+k}$) when $n>1$ (see [Ki]). The
``expected'' dimension of $\hol{A}(C, V)$ (where $A$ is a fixed
homology class in $H_2(V)$ and $f_*[C]=A$ for all $f\in\hol{A}$ can
however be computed and at generic smooth points it is given by $c_1.A
+ n(1-g)$. In our case $c_1=n+1$ ($V=\bbp^n$) and $A$ is $k$ fold the
generator in $H_2(V)$.

\vskip 10pt
\noindent{\bf\S3.1 The configuration space model for $\bmap{}(C,\bbp^n)$}

In 1.7 we pointed out to the existence of a model for $\map{}(C,
\bbp^n )$ in terms of bounded multiplicity symmetric products
$\sp{k}_n(-)$ defined in \S2. More
precisely, consider $C-U$ where $U$ is a little open disc around the
basepoint $*$ ($C-*$ isotopy retracts onto $C-U$). Let $U_k$ be a
nested sequence of neighborhood retracting onto $*$, then one defines
the space $\spy_n(C)$ as the direct limit of inclusions
$\sp{k}_n(C-U_k)\lrar\sp{k+1}_n(C-U_{k+1})$, $D\mapsto D+x_k, x_k\in
U_{k}-U_{k+1}$ (here $\{x_k\}$ is a sequence of distinct points
converging to $*$).

One can map $\spy_n(C-*)$ to the based mapping space by {\sl
scanning}
$$S: \spy_n(C-*)\lrar\bmap{0}(C,\bbp^{n})\leqno{3.4}$$ 
This map is given as follows (details in [K2]): 
identify canonically each neighborhood
$D(x)$ of $x\in C-*$ with a closed disc $D^2$ (this is possible
because $C-*$ is parallelizable). It follows that to every
configuration in $\spy_n(C-*)$ one can restrict to $D(x)$ and see
a configuration in $\spy_n(D(x),\partial D(x))=\spy_n(D^2,\partial
D^2)$ (the relative construction means that configurations are
discarded when they get to the boundary). It can be checked that 
$\spy_n(D^2,\partial D^2)=\sp{n}(S^2)=\bbp^n$ and hence 3.4.
It turns out that the map $S$ in 3.4. is a homotopy equivalence when
$n>1$ or a homology equivalence when $n=1$ (cf. [K2]). 
 
\noindent{\bf Lemma} 3.5:~{\sl Assume $g\geq 1$, $k\geq n\geq 1$
and $C'=C-*$.  Then the collar inclusion
$\sp{k}_n(C')\rightarrow\spy (C')$ induces an epimorphism  
on $H_1$ (for $n>1$ it is in fact an isomorphism).}

\noindent{\sc Proof}: A cohomology class in $H^i(C',\bbz )$ is
represented by (the homotopy class) of 
a map $C'\lrar K(\bbz, i )$ and since the target is an
abelian topological group, this map extends multiplicatively to
$$C'\fract{i_1}{\lrar}\sp{k}_n(C')\fract{i_2}{\lrar}
\sp{k}(C')\lrar K(\bbz, i)$$
and hence gives rise to a class in $H^i(\sp{k}_n(C'))$.
The ``inclusion'' $i_1:C'\rightarrow\sp{k}_n(C')$ (constructed earlier) 
is then surjective in cohomology and hence injective
in homology.  Since $\pi_1(\sp{k}(C'))=H_1(\sp{k}(C'),\bbz
)=H_1(C';\bbz )$, the composite $i_2\circ i_1$ must be an isomorphism
on $H_1$. This then implies that $H_1(\sp{k}_n(C'))\lrar
H_1(\sp{k}(C'))$ is necessarily surjective. It can be checked that as
soon as $n\geq 2$, $\pi_1(\sp{k}_n(C'))$ abelianizes ([K3]) and from
there that $H_1(\sp{k}_n(C'))\cong
H_1(\sp{k}(C'))=\bbz^{2g}$.\hfill\za

\noindent{\bf Lemma} 3.6:~{\sl Assume $g\geq 1$. The map
$\bmap{k}(C,\bbp^n)\fract{\alpha}{\lrar}
\bmap{k}(C,\bbp^{\infty})\simeq (S^1)^{2g}$, induced from
post-composition with the inclusion
$\bbp^n\hookrightarrow\bbp^{\infty}$, is an isomorphism at the level
of $H_1$ when $n>1$ and a surjection when $n=1$.}

\noindent{\sc Proof}: Consider the following homotopy diagram
$$\begin{array}{ccc}
  \spy_n(C-*)&\fract{\subset}{\ra 3}&\spy(C-*)\\
  \decdnar{}&&\decdnar{\simeq}\\
  \bmap{0}(C,\bbp^{n})&\fract{\alpha}{\ra 3}&
  \bmap{0}(C,\bbp^{\infty})
\end{array}
$$
The top map is an isomorphism on $H_1$ (and only a surjection when
$n=1$) according to the previous lemma.  Since both vertical maps are
isomorphisms on $H_1$ as well (by 3.4), the claim follows.  \hfill\za

\vskip 10pt
\noindent{\bf\Large\S4 A Spectral Sequence of Milgram}
\vskip 10pt

The following spectral sequence appears in special cases in [BCM],
[B], [K1] and [KM] and the main ideas trace back to [M1].  Let $X$
be a space with basepoint $*$ and let $E(X)\hookrightarrow \spy
(\bigvee^k X)$ be a submonoid of $\spy (\bigvee^k X) = \prod_k\spy
(X)$.  Given a map $X\lrar E(X)$, it can always be extended to a map
$\nu_0: \sp{r}(X)\lrar E(X)$ (additively) and then to a map
$$\nu: \coprod_{r\geq 1}\sp{r}(X)\times E \lrar E, \ \ \nu (x,y) =
\nu_0(x) + y$$
Our interest is to study the complement $Par(X) = E(X) -
\hbox{Im}(\nu)$ (or {\sl Particle} space). We reserve the notation
$C(X)$ for the standard configuration space of distinct points (see
4.1 below).  Such generalized families of configuration spaces are
studied in [K2].

\noindent{\sc Examples} 4.1:\hfill\break
$\bullet$ (i) Let $k=1$ and consider the map $M\lrar E(M) =\spy (M)$,
$x\mapsto 2x$. Points in Im($\nu$) are finite sums $\sum n_ix_i$ where
$n_i\geq 2$ for at least one index $i$. In this case
$$Par(M) = C(M) = \{\sum x_i, x_i\neq x_j~\hbox{for}~i\neq j\}$$ 
Similarly we can map $M\lrar E$ by sending $x\mapsto (n+1)x$ and in
this case the discriminant space is the space $\spy_n(M)$ described
in 3.4.
\hfill\break
$\bullet$ (ii) The divisor space $\d{}^n(M)$ introduced in \S3 is the particle
space obtained as the complement of the diagonal map
$$M\lrar\prod^n\spy (M)=E(M), ~x\mapsto (x,x,\ldots, x)$$
$\bullet$ (iii) If $\mu: M\lrar G$ is a map into an abelian group $G$, then
$\mu$ can be extended to $\spy (M)$ and one can define
$$E(M,\mu ) = \{(D_1,\ldots, D_k)\in \prod_k\spy (M)~|~\mu (D_i)=\mu (D_j)~
\hbox{for all}~i, j\}$$
This is a submonoid since by construction $\mu (D + D') = \mu (D)+\mu (D')$.
The main example we study in this paper is when $M=C$ is a curve
and $\mu$ is its Abel-Jacobi map. 

Consider the following construction
$$DE(M) = E\times_{\nu}\spy (cM)$$
where $cM$ is the cone on $M$ and where the twisted product
$\times_{\nu}$ means the identification via $\nu$ of the points
$$({\vec\zeta}, (t_1,z_1), \ldots , (t_l,z_l))
\sim ({\vec\zeta}+ \nu (z_i), 
(t_1,z_1)\ldots {\widehat{(t_i,z_i)}}\ldots
(t_l,z_l))$$
whenever $t_i=0$ is at the base of the cone (here {\it hat} means deletion).
The identification above ``cones off'' the image of ${\nu}$
in $E$ and so we expect that $H_*(DE) = H_*(E/Im\nu )$. Write
$$E_{n_1,\ldots, n_k} = E(M)\cap \left(\sp{n_1}(M)\times\cdots\times
\sp{n_k}(M)\right)$$
and suppose $\nu: M\lrar E(M)$ lands in $E_{l_1,\ldots, l_k}, l_j\geq
1$. We can then consider the filtration of $DE$ by subspaces
$$DE_{k_0,k_1,\ldots, k_n}(C_g) =
\bigcup_{{{i_j+ll_j\leq k_j}\atop 1\leq j\leq k}}
E_{i_0,i_1,\ldots, i_n}\times_{\nu}\sp{l} (cC_g)$$
There are (well-defined) projection maps
$$p_{k_0,k_1,\ldots, k_n}\colon
DE_{k_0,k_1,\ldots, k_n}\lrar E_{k_0,k_1,\ldots, k_n}/
\left\{\hbox{Image}(\nu)\right\}$$
sending $(v_1, \dots, v_s, (t_1, w_1), \dots, (t_r, w_r))$ to
$(v_1, \dots, v_s)$.
These maps are acyclic and hence induce
isomorphisms in homology (the proof in the case of the standard
configuration spaces 4.1 (i) is given in [BCM], lemma 4.6 and relies
mainly on the fact that $\sp{l}(cX)$ is contractible for every $l$).

Let's assume $l_j=d, 1\leq j\leq k$ and write $(D)E_{i,\ldots, i}=
(D)E_{i}$. Suppose $Par_i(M)= E_{i} - Im(\nu )\cap E_{i}$ is an oriented
manifold of dimension $N(i)$ (it will be for all cases we consider
here otherwise we can use $\bbz_2$ coefficients).  
Then by Alexander-Poincar\'e duality we have an isomorphism
$\tilde H^{N(i)-*}(Par_{i}(M); \bba )\cong H_*(E_{i}/Im(\nu );
{\bba } )$ for commutative rings $\bba$.
Combining this with the previous paragraph we get

\noindent{\bf Proposition} 4.2: {\sl~There are isomorphisms
\begin{eqnarray*}
\tilde H^{N(i)-*}(Par_{i}(M); {\bba })&\cong& H_*(DE_{i}; {\bba } )\\
\tilde H^{N(i)-*}(Par_{i}(M-*); {\bba } )&\cong& H_*(DE_i/
\bigcup_i DE_{i,\ldots, i-1,\ldots,i};{\bba } )
\end{eqnarray*}}

Let $LE_i$ be the quotient $E_{i,\ldots, i}/\bigcup E_{i,\ldots,
i-1,\ldots, i}, i\geq 1$. The map $\nu$ is made out of maps
$$\nu:\sp{r}(M)\times E_i\lrar E_{i+rd}$$
and so we can consider the quotient model
$$QE_k = DE_{k,\ldots, k}/
\bigcup DE_{k,\ldots, k-1,\ldots, k}=
\bigcup_{i+dj=k} LE_i\times_{\nu} (\sp{j}(cM)/\sp{j-1}(cM))$$
One can filter this complex by $j$ (in which case one obtains an
analog of the Eilenberg-Moore spectral sequence with $E^2$ term
a Tor term; see [K1], [BCM]) or one can filter according to $i$ by 
letting
$${\cal F}_r = \bigcup_
{i\geq r\atop {i+dj=r}} LE_i\times_{\nu} (\sp{j}(cM)/\sp{j-1}(cM))$$
In this case we get the spectral sequence

\noindent{\bf Theorem} 4.3:~{\sl There is a spectral sequence converging
to $H^{N(k)-*}Par_k(M-*)$ with $E^1$ term
$$E^1 = \coprod_{i+dj=k\atop i\geq 1} H_*(LE_i;\bba )\tensor H_*(\sp{j}(\Sigma
M),\sp{j-1}(\Sigma M);\bba )$$
and where the $d^r$ differentials are obtained from a chain
approximation of the identification maps
$\nu:\sp{j}(M)\times LE_i\lrar LE_{i+d}$, $(x,y)\lrar \nu (x) + y$.}

\noindent{\sc Sketch of proof}:  (see [K1] and [KM] for details)
Consider ${\cal F}_r$ as described above. Then a chain complex for
${\cal F}_r$ is given as
$$\bigoplus C_*(LE_i)\otimes_{\nu} C_*(\sp{l}(cM)/\sp{l-1}(cM))$$ 
Here the symmetric product pairing as well as the identification given
by $t$ can be chosen to be cellular.  We need determine the homology
of this complex.
An interesting fact is that one can construct chain complexes
$$C_*(\spy (cM)) = C_*(\spy (M))\otimes C_*(\spy (\Sigma M))
\leqno{4.4 }$$
where the identification above is given as a bigraded differential
algebra isomorphism.  This follows from the fact (cf.[M1])) that
$C_*(\spy (c M))$ can be identified with the acyclic bar construction
on $C_*(\spy (M)$ and that $C_*(\spy (\Sigma M))$ can be identified
with the reduced bar construction on $C_*(\spy (M)$ .
Therefore the boundary on cells of ${\cal F}_r$ can be made
{\sl explicit}.  First a cell in $C_*(\sp{l}(cM)/\sp{l-1}(cM))$ can be
written as $$c_*\otimes |a_1|\cdots |a_l|$$ 
(according to the decomposition 4.4) and the boundary decomposes into
$$\partial = \partial c_*\otimes |a_1|\cdots |a_l| + \nu_*(c_*\otimes
a_1)\otimes |a_2|\cdots |a_l| + c_*\otimes \partial_B(|a_1|\cdots
|a_l|)\leqno{4.5}$$ 
where $\partial_B$ is the usual bar differential.  The induced
boundary on ${\cal F}_r/{\cal F}_{r+1}$ (which describes $d_0$) is
given by the first and last term and the homology of the complex
$\partial: C_*({\cal F}_r/{\cal F}_{r+1}) \rightarrow C_*({\cal
F}_{r+1}/{\cal F}_{r+2})$ is given by the expression in 4.3. Remains
to identify the $d^r$ differentials and these are deduced by the
middle term in 4.5 according to the filtration term in which they land.
\hfill\za

\vskip 10pt
\noindent{\bf\S4.1 An application: Configurations with bounded multiplicity}

The homology of $\spy_n (C'), C'=C-*$ can be calculated from the
filtration pieces $\sp{k}_n(C-*)$ as follows.  Start by observing that
the configuration space $\spy_n(C)$ is the discrimant set in $\spy
(C-*)$ of the image of $\nu: \spy (C)\lrar \spy (C)$ which is
multiplication by $n+1$ (as described in 4.1 (i)). In this particular
case 4.3 takes the form (with field coefficients)

\noindent (4.6 )\ \ {\sl There is a spectral sequence converging to  
$H_{2k-*}(\sp{k}_n(C-*);\bbf )$, $n\geq 1$, with $E^1_{i,j}$ term
(to which we refer as $E^1(k)$ )
$$\bigoplus_{i+(n+1)j=k\atop r+s=*} 
H_r(\sp{i}(C),\sp{i-1}(C);\bbf )\tensor H_s(\sp{j}(\Sigma
C_g),\sp{j-1}(\Sigma C_g);\bbf ).$$}

To see this, one simply observes that $\sp{k}_n(C-*)$ is open in
$\sp{k}(C-*)$ which is a $k$ dimensional complex manifold and then one
applies 4.3.  We point out that such a spectral sequence was
considered in the case of $n=1$ in ([BCM], theorem 4.1) and for $g=0$
and all $n$ in [K3]. The differentials as explained in the proof of
4.3 are induced from a cellular approximation of the maps
$$\nu: \sp{j} (C)\lrar\sp{(n+1)j} (C),~~x\mapsto (n+1)x\leqno{4.7}$$ 
More explicitly in this case, a chain complex for a Riemann surface
$C$ of genus $g$ is given by $2g$ one dimensional classes (which we
label $e_1,\ldots, e_{2g}$) and a two dimensional orientation class
$a=[C]$. Now the homology of $\spy (C)$ is generated as a ring
by the symmetric products of these classes; i.e.
$$H_*(\spy(C)) = E (e_1,\ldots, e_{2g})\tensor\Gamma [a]$$
and $H_*(\sp{n}(C))$ consists of all $n$-term products in the complex
above (here $E$ is an exterior algebra and $\Gamma$ is divided power
algebra; see [K1] for details). It turns out that this homology
{\bf embeds} in $C_*(C)$ and so one can think of
the product of these classes as cells as well. We can investigate the
boundary term 4.5 on these classes. The map $\nu$ is given by
the composite
$C\fract{\Delta}{\lrar}C^{\times n}\rightarrow\sp{n}(C)$.
The primitive classes $e_i$ map to $\sum 1\tensor\cdots\tensor
e_i\tensor\cdots\tensor 1$ and hence map into $H_*(C)\subset
H_*(\sp{n+1}(C))$. For $n>1$ they clearly vanish in 
$H_*(\sp{n+1}(C),\sp{n}(C))$
and so are not seen in the spectral sequence.
The class $a$ on the other hand maps via $\Delta$ to the class
$\sum 1\tensor \cdots \tensor 1\tensor C\tensor 1\tensor\cdots 1 + 
\sum 1\tensor \cdots e_i\tensor \cdots \tensor e_j\tensor
\cdots\tensor 1$ in $H_2(C^{n+1})$. The projection into
$H_*(\sp{n+1}(C),\sp{n}(C))$ vanishes
if $n>1$ and is non trivial if $n=1$. More explicitly we have

\noindent{\bf Lemma} 4.8: 
{\sl When $n>1$, all $d^r$ differentials vanish and the spectral
sequence above collapses at the $E^1$ term. When $n=1$, there are 
higher differentials generated by $d^1(1\tensor |a|) = 
2\sum e_ie_{i+g}$, $1\leq i\leq g$.}

\noindent{\sc Example} 4.9 ($H_1(\sp{k}(C-*))$): 
There are $2g$ one dimensional (torsion
free) classes ${\tilde e}_i$ in $H_1(\sp{k}_n(C-*))$ corresponding to
the classes $e_ia^{k-1}\in E^1_{k,0}$ (which have dimension
$2(k-1)+1$). Also and when $n=1$, 
the class $a^{k-2}|a|$ in $E^1$ gives rise to a generator
in $H_1(\sp{k}_1(C-*))$ for all $k>1$. But
$\partial |a|=2\sum e_ie_{i+g}$ (according to 4.8) 
and hence this is a 2-torsion class; i.e.
$$H_1(\sp{k}_n(C-*);\bbz )
\cong \left\{{\bbz^{2g},\ \ \ \ \hbox{when}\  n>1, k\geq 1}\atop
{\bbz_2\oplus\bbz^{2g},\ \hbox{when}\  k> n=1}\right.$$
This also corresponds to $H_1(\bmap{0}(C,\bbp^n ))$ as will be clear
shortly (and as is expected according to 1.4!).

\noindent{\sc Example} 4.10: There are inclusions
$H_*(\sp{k}_n(C'))\hookrightarrow H_*(\sp{k+1}_n(C'))$ induced by 
a map of $E^1$ terms (in 4.6 above)
$$ x\tensor y\mapsto xa\tensor y$$
As a corollary one can easily deduce the following standard fact
(for another argument, see [K3] this volume): 
{\sl There are homology splittings (here $\sp{0}(C')=*$) 
$$H_*(\sp{k}_n(C'))\cong \bigoplus_{1\leq i\leq k}
H_*(\sp{i}_n(C'),\sp{i-1}_n(C'))$$}

\vskip 10pt
\noindent{\bf\S4.2 The homology of $\bmap{0}(C,\bbp^n )$}

We now determine $H_*(\bmap{}(C,\bbp^n);\bbf )$ for $\bbf=\bbz_2,
\bbz_p$ as a straightforward application of the calculations in \S4.1.
Recall that when $g>0$, the surface $C_g$ is described topologically
by a 2-disc attached to a bouquet of $2g$ circles via the mapping
which wraps around as a product of commutators. More precisely one has
a cofibration sequence
$$S^1\fract{f}{\ra 2}\bigvee^{2g}S^1\fract{i}{\hookrightarrow} C_g
\fract{\pi}{\lrar} S^2\leqno{4.11}$$
where $i:\bigvee^{2g}S^1\hookrightarrow C_g$
is the one skeleton inclusion, and the map
$f$ is given as a product of commutators
$[x_1,x_2][x_3,x_4]\cdots [x_{2g-1},x_{2g}]$ (with the $x_i$'s denoting
the generators of $\pi_1(\bigvee_{2g} S^1) = \bbz^{*2g}$).
Applying the $\bmap{}(-,\bbp^n)$ functor to 4.11 yields the fibration
$$\Omega^2(\bbp^n)\lrar\bmap{}(C_g,\bbp^n)\lrar\Omega(\bbp^n)^{2g}
\leqno{4.12}$$
We have $H^*(\Omega(\bbp^n)^{2g})=H^*(S^1)^{\tensor 2g}\tensor
H^*(\Omega^2S^{2n+1})$. Note that 4.12 has simple
coefficients (since the fiber is a loop space).  Write again (cf. \S2)
$H^*(\Omega^2S^3,\bbz_p )$ as an exterior $E (x_1,\ldots,
x_{2p^{i+1}-1},\ldots )$ tensor a truncated algebra
$P_T(y_{2p-2},\ldots, y_{2p^i-2},\ldots )$ where the $x$'s and $y$'s
are generators in the stated dimensions.

\noindent{\bf Theorem} 4.13:~{\sl Assume $g>1$. Then the Serre
spectral sequence for 4.12 collapses at $E^2$ when $n>1$.  When $n=1$,
the spectral sequence collapses with $\bbf_2$ coefficients but has the
mod-$p$ differentials ($p>2$)
\begin{eqnarray*}
d_{p^i}(x_{2p^i-1})&=& {1\over p^i}
\left(\sum_1^gf_{2i-1}f_{2i}\right)^{p^i}\\
d_{p^i}(y_{2p^{i+1}-2})&=&\left[{1\over p^i}
\left(\sum_1^gf_{2i-1}f_{2i}\right)^{(p-1)p^i}\right]x_{2p^i-1}.
\end{eqnarray*}
where the $f_i$ are the one dimensional generators of
$\Omega(\bbp^1)^{2g}\simeq (S^1)^{2g}\times (\Omega S^3)^{2g}$.}

\noindent{\sc Proof}: This is a counting argument.
The homology splitting in 4.10 holds for
$k=\infty$ and combining this with 1.7 (and field coefficients) we see
that
$$H_*(\bmap{k}(C,\bbp^n ))\cong \bigoplus_{k\geq 1}
H_*(\sp{k}_n(C'),\sp{k-1}_n(C'))$$
We simply need identify generators. From 4.6, we read off
the correspondence between generators in $E^{\infty}$ and generators
in the serre spectral sequence as follows
\begin{eqnarray*}
e_i&\mapsto& f_i\ \ \ \ \hbox{and}\ \ \ |a|\mapsto x_1\\
|e_i|'s &\mapsto&\hbox{generators of}\ H^2(\Omega S^3)^{2g}
\end{eqnarray*}
($|e_i|, |a|\in E(2)$ and $e_i\in E(1)$, but they
propagate to $E(k)$ after multiplying by suitable powers of $a$
as indicated in 4.10).
Now $a$ generates a divided power algebra in $H_*(\spy (C))$
and hence $|a|$ generates an $H_*(K(\bbz, 3))$ in $H_*(\spy (\Sigma C))$
(as is known, the groups $H_*(K(\bbz, 3))$ and $H^*(\Omega^2S^3)$ are formally
``dual'' to each other). More precisely, we can construct a
correspondence between the two as follows (with mod-$p$ coefficients):
let $\gamma_i$ be the divided power generators in $\Gamma (a )$, then
$$ |\gamma_{p^i}|\mapsto x_i,\ \ \
|\gamma_{p^i}^{p-1}|\gamma_p|\mapsto y_{i}$$
The rest of the proof is now straightforward as the non-zero
differentials for $p>2$ are entirely generated by the one in 4.8
(compare [K1]).\hfill\za

\vskip 10pt
\noindent{\bf\Large\S5 Segal's Stability Theorem}
\vskip 10pt

In this section we derive Segal's result based on the spectral
sequence in 4.3 which in this particular context takes the form

\noindent (5.1)\ \ {\sl There is a spectral sequence converging to
$H^{*}(\hol{k}(C_g,\bbp^n ))$, $k\geq 2g-1$ with $E^1$ term
$${_h}E^1 = \bigoplus_{i+j=k\atop i\geq 1} E^1_{i,j} = 
\bigoplus_{i+j=k\atop i\geq 1} H_{*'}(LE_i;\bba )\tensor 
H_{*''}(\sp{j}(\Sigma
C_g),\sp{j-1}(\Sigma C_g);\bba ).$$
and identifiable $d^1$ differential. 
Here $* =2(k-g)(n+1)+2g-*'-*''$.}

The terms $LE_i$ are constructed out of the Jacobi variety $J$ and its
stratifications. Recall that by construction $E_i$ is the pull-back of
the diagonal
$$\begin{array}{ccc}
  E_{i}&\fract{}{\ra 2}&\sp{i}(C_g)\times\cdots\times\sp{i}(C_g)\\
  \decdnar{}&&\decdnar{\underbrace{\mu\times\cdots\times \mu}_{n+1}}\\
  J&\fract{\Delta}{\ra 2}&J(C_g)\times\cdots\times J(C_g)
\end{array}
\leqno{5.2}$$
The image under $\mu$ of $\sp{i}(C)$ in $J(C)$
is denoted by $W_i$. $W_i$ has complex dimensions $i$ and for
for $i\geq g$, $\mu$ is surjective and $W_i=J$. 

We nee recall at this stage that the Abel-Jacobi map
$\mu:\sp{i}(C_g)\rightarrow J(C_g)$ is an analytic fibration in the
``stable" range $i\geq 2g-1$ with fiber $\bbp^{i-g}$ (that the
dimension of the fiber stabilizes is a consequence of
Riemann-Roch. That $\mu$ is actually a fibration is a theorem of
Mattuck [Ma]).  So from 5.2 it is easy to deduce that for $i\geq 2g$ 
$$H_*(LE_i,\bba )\cong H_*(J(C_g),\bba )\tensor
H_*(S^{2(i-g)(n+1)};\bba) \leqno{5.3}$$
(i.e. $LE_i$ is obtained by first pulling back
$\sp{i}(C)^{n+1}$ over $J(C)$ via $\Delta$ and then
collapsing the fat wedge in the fiber $(\bbp^{n-i})^{n+1}$.)

\noindent{\sc Example} 5.4: (the torsion free one-dimensional
classes). Note that when $i\geq 2g$, we deduce from 5.3 that there are
$2g$ classes in $H_*(LE_k)$ of dimension $2(k-g)(n+1) + 2g-1$ which
yield $2g$ torsion free generators in $H_1(\bhol{k})$. 

\vskip 10pt \noindent{\sc Stable Classes and Segal's theorem}: When
$i\geq 2g-1$ (the {\sl stable range}), the terms in ${_h}E^1_{i,j}$
that survive to ${_h}E^{\infty}$ yield `stable' classes in
$H^*(\bhol{})$ (and hence in $H_*(\bhol{})$). The following is shown
in ([KM], lemma 8.3)

\noindent{\bf Proposition} 5.5:~{\sl The induced map
$H_*(\bhol{k}(C_g,\bbp^n))\lrar H_*(\bmap{k}(C_g,\bbp^n ))$
is an isomorphism on the stable classes}.

\noindent{\sc Proof} (sketch of alternate argument): When $i\geq
2g-1$, $E_i$ is an $(i-g)(n+1)+2g$ complex and maps to
$\sp{(i-g)(n+1)+2g}(C)$ (via a factoring of the map $E_i\rightarrow
\sp{i}(C)^{n+1}\fract{+}{\lrar}\sp{i(n+1)}(C)$ (where $+$ is
concatenation). It follows that the stable terms in ${_h}E^1$
(corresponding to $i\geq 2g-1$) map to $E^1(N)$ (see 4.6; here
$N=(k-g)(n+1)+2g)$). The main point is that in the stable range the
spectral sequence in 5.1 and the one in 4.6 behave identically (same
differentials described as in 4.8 above and [KM], lemma 7.8). 
The stable terms surviving to ${_h}E^{\infty}$ therefore map
isomorphically to $E^{\infty}(N)$ and so the dual ``stable classes''
in $H_*(\bhol{k})$ have isomorphic image in $H_*(\sp{k}_n(C-*))$;
i.e. in $H_*(\spy_n(C-*))\cong H_*(\bmap{0}(C,\bbp^n )$. Details
omitted.\hfill\za

The unstable terms which appear when $i < 2g-1$ in the spectral
sequence above form the interesting part and are harder to track down
for their existence depends strongly on the geometry of the curve. It
follows however that their contribution only starts appearing above a
certain range and hence up to that range the homology of
$\hol{k}(C_g,\bbp^n)$ only consists of stable classes and by 5.5
must coincide with that of $\map{k}(C_g,\bbp^n ))$.  This is exactly
the essence of the stability theorem of Segal and we fill in the
details below.

\noindent{\bf Proposition} 5.6:~$H_*(LE_i)=0$ for $*> i(n+1)+2g$ and
for all $1\leq i\leq 2g$.

\noindent{\sc Proof:} Define
$$W_i^r= \{x\in\mu (\sp{i}(C_g))~|~\mu^{-1}(x)=\bbp^m, m\geq r\}$$
A well-known theorem of
Clifford asserts that in the range $1\leq i\leq 2g$ the maximum of $r$
is $i/2$; that is $W_i^{r}=\emptyset$ for $r> {i\over 2}$.
One can then filter 
$\mu (\sp{i}(C)) = W_i$ by the descending filtration
$$W_i^{i\over 2}\subset \cdots\subset W_i^1\subset W_i$$
This leads to a spectral sequence converging to
$H_*(LE_i)$ with $E^1$ term
$$\sum_r^{i\over 2}H_*(W_i^{r}, W_i^{r+1})\tensor H_*(S^{2r(n+1)})$$
Since the $W_i^r-W_i^{r+1}$ are algebraic subvarieties of $J(C_g)$ ([ACGH])
we must have that $H_*(W_i^r,W_i^{r+1} )=0$ for $*>2g$. 
This means that each term in the $E^1$ term above 
has no homology beyond $2{i\over 2}(n+1)+2g$ and
hence
$$H_*(LE_i)=0~~\hbox{for}~~*>2{i\over 2}(n+1)+2g$$
as asserted.  Note that when $i=2g$, we have that $m$ is uniformly
$i-g=g ={i\over 2}$ and the proposition is still valid in this case.
\hfill\za

\noindent{\bf Corollary} 5.7: (Segal)~{\sl The inclusion 
$$\hol{k}(C_g,\bbp^n)\hookrightarrow\map{k}(C_g,\bbp^n )$$
is a homology isomorphism up to dimension $(k-2g)(2n-1)$}

\noindent{\sc Proof:}
Look in 4.3 at the unstable terms given by 
$$H_*(LE_i)\tensor H_{*'}(\sp{j}(\Sigma C_g),\sp{j-1}(\Sigma C_g)),~
i\leq 2g-1$$
According to 5.6 these terms vanish for $*> i(n+1)+2g$.
By duality, they contribute therefore no homology to 
$H_*(\hol{k}(C_g,\bbp^n );\bba )$ for
$$*\leq \left[2(k-g)n +2k\right]- \left[i(n+1)+2g+3(k-i)\right]$$
The expression on the left attains its minimum when $i=2g$ that is when
$$\left[2(k-g)n+2k\right]- \left[2g(n+1)+2g+3(k-2g)\right]
= (k-2g)(2n-1).$$
So when $*<(k-2g)(2n+1)$ the unstable terms do not 
contribute to $H_*(\hol{k=i+j}(C_g,\bbp^n );\bba )$ and the proof 
now follows from proposition 5.5.
\hfill\za

\vskip 10pt 
\noindent{\bf\Large\S6 The Fundamental Group}
\vskip 10pt

Since the attaching map of the two cell of $C_g$ is given by a product
of commutators corresponding to Whitehead products (see 4.11), its
suspension must be null-homotopic. This implies that $\Sigma C_g$
splits as a wedge $\Sigma C_g\simeq\bigvee S^2\vee S^3$ and one has
$$\displaystyle \Sigma^i C_g\simeq\bigvee_1^{2g} S^{i+1}\vee
S^{i+2},~~i\geq 1.\leqno{6.1}$$ 

\noindent{\bf Lemma} 6.2:~{\sl All components of $\bmap{}(C_g,X)$ are
homotopy equivalent, and $\forall i>1$, there is an isomorphism of
homotopy groups
$$\pi_i(\bmap{0}(C_g,X))=\bigoplus^{2g}\pi_{i+1}(X)^{2g}
\oplus\pi_{i+2}(X).$$}

\noindent{\sc Proof:} The statement about components is standard
and we denote by $\bmap{0}(C_g,X)$ the component of null-homotopic maps.
On the other hand and by definition 
$\pi_i(\bmap{0}(C_g,X)) = [S^i\wedge C_g,X]_*$.
The splitting 6.1 yields ({\sl as sets}) the bijection 
$$[S^i\wedge C_g,X]_* = [\bigvee_{2g} S^{i+1}\vee S^{i+2},X]_* 
=\pi_{i+2}(X)\oplus\bigoplus^{2g}\pi_{i+1}(X)$$
In order for the decomposition above to induce a group homomorphism,
it is necessary that 6.1 be a decomposition as ``co-H spaces", that
is that there is a map 
$f_i: \Sigma^iC_g\fract{f_i}{\ra 2}\bigvee_1^{2g} S^{i+1}\vee S^{i+2}$
which commutes with the pinch maps up to homotopy.  When $i\geq 2$,
$f_i=\Sigma f_{i-1}$ is a suspension and hence is automatically a co-H
map. The claim follows.  \hfill\za

When $i=1$, the decomposition in 6.2 is not necessarily 
a group decomposition. The long
exact sequence in homotopy associated to $\Omega^2X\rightarrow
\bmap{}(C_g,X)\rightarrow (\Omega X)^{2g}$ together with 6.2 indicate
however that there is a short exact sequence of abelian groups
$$0\lrar\pi_{3}(X)\lrar\pi_{1}(\bmap{0}(C_g,X))
\lrar\pi_{2}(X)^{2g}\lrar 0$$
When $X=\bbp^n$, $\pi_2(\bbp^n )
\cong \bbz$ and $\pi_3(\bbp^n ) \cong\pi_3(S^{2n+1})$ which is zero
when $n>1$. This shows that
$$\pi_1\bmap{0}(C,\bbp^n ) =\bbz^{2g},~~\hbox{when}~~n>1\leqno{6.3(a)}$$
When $n=1$, there is a short exact sequence
$$0\lrar\bbz\lrar\pi_1(\bmap{0}(C,\bbp^1 ))\lrar\bbz^{2g}\lrar 0
\leqno{6.3(b)}$$
This is a central extension which turns out to be non-trivial as we
now show.

First observe that any Riemann surface has meromorphic functions for
sufficiently high degrees $k$ (this is certainly true when $k\geq 2g$
and we will assume this is the case throughout the section). Let $f$
be any such function and construct the map
$\rat{1}(\bbp^1)\lrar\bhol{k}(C, \bbp^1)$ by post-composition with
$f$. $\rat{1}(\bbp^1)$ is the set of pairs of distinct points (a root
and a pole) in $\bbc$ and has the homotopy type of $S^1$.

\noindent{\bf Lemma} 6.4:~{\sl The inclusion $f^{!}:
\rat{1}(\bbp^1)\lrar\bhol{k}(C, \bbp^1)$ induces an injection at the
level of $\pi_1$.}

\noindent{\sc Proof}: 
Observe that the diagram below commutes strictly
$$\begin{array}{ccc}
  \rat{1}(\bbp^1)&\fract{f^{!}}{\lrar}&\bhol{k}(C, \bbp^1)\\
  \decdnar{i_1}&&\decdnar{i_2}\\
  \Omega^2_1S^2&\fract{g}{\lrar}&\bmap{k}(C, S^2)
\end{array}
\leqno{6.5}$$
The vertical maps are inclusions.
The bottom map on $\pi_1$ is an injection according to 6.3(b) and hence
$g\circ i_1$ is also and injection on $\pi_1$ (since $i_1$ on $\pi_1$
is an isomorphism between two copies of $\bbz$). The composition
$i_2\circ f^{!}$ is therefore injective on $\pi_1$ and hence so
is $f^{!}$ as desired.
\hfill\za

Next and from the description of $\bhol{k}$ 
and $\d{k,k}(C-*)=\sp{k}(C-*)^2-\Delta$ where $\Delta$ is the
generalized diagonal consisting of pairs of divisors with (at least)
one point in common (see \S3), we have the pullback diagram
$$\begin{array}{cccc}
  \bhol{k}(C,\bbp^1)&\fract{\phi}{\lrar}&\d{k,k}(C-*)\\
  \decdnar{\pi}&&\decdnar{\mu^2}\\
  J(C)&\fract{\Delta}{\lrar}&J(C)^2
\end{array}
\leqno{6.6}$$
where $\pi:
\bhol{k}(C,\bbp^n)\fract{p}{\lrar}\sp{k} (C-*)\fract{\mu}{\lrar} J(C)$
is the map that sends a holomorphic map to the equivalence class of
its divisor of zeros and $\Delta$ is the diagonal. 
The top horizontal map $\phi$ is an inclusion. We denote by $F$ the
homotopy fiber of $\mu^2$ and $\pi$.

\noindent{\bf Lemma} 6.7:~{\sl The map $p: \bhol{k}(C,\bbp^1 )\lrar\sp{k}
(C-*)$ is surjective at the level of fundamental groups}.

\noindent{\sc Proof}: For $k\geq 2g-1$ the restriction of the Mattuck
fibration (cf. \S5) to $\sp{k}(C-*)$ becomes a complex vector bundle
$$\bbc^{k-g}\lrar\sp{k}(C-*)\fract{\mu}{\lrar} J(C)$$
(i.e. the set of linearly equivalent divisors avoiding a point is
a hyperplane in $\bbp^{k-g}$).
It follows that $\mu_*$ is an isomorphism on $\pi_1(\sp{k}(C-*))$. Since the
fiber of $\pi$ is connected (see 6.10), it follows that $\pi$ is
surjective at the level of $\pi_1$ and hence is $p$.  \hfill\za

\noindent{\bf Theorem} 6.8:~{\sl Suppose $k\geq 2g$. Then
$\pi_1(\bhol{k}(C,\bbp^1 ))$ is generated (multiplicatively) by
classes $e_1,\ldots, e_{2g}$ and $\tau$ such that the commutators
$$[e_i, e_{g+i}] = \tau^2$$
and all other commutators are zero. }

We will prove this in a series of lemmas. To begin with, lemmas 6.4
and 6.7 show that there is a sequence
$$0\lrar\bbz\fract{f_* }{\lrar}\pi_1\bhol{k}(C,\bbp^1 )
\fract{p_*}{\lrar}\bbz^{2g}\lrar 0\leqno{6.9}$$
which is exact at both ends. In 6.13 we show that this sequence maps
to 6.3(b) and hence Im($f_* )\subset$ Ker
$(p_*)$.  That Ker$(p)\subset$Im$(f_* )$ follows from 6.11 and the
discussion next.

\noindent{\bf Lemma} 6.10:~{\sl $\pi_1(\d{k,k}(C-*)), k>1$ has
torsion free generators $a_i, b_j, 1\leq i,j\leq 2g$ and $\tau$. The
$a_i$ (resp. $b_i$) are represented by a root (resp. a pole) going around
the generators in a symplectic basis of $H_1(C-*)$, while $\tau$ is
given by a zero (or pole) going around a pole (resp. a zero).}

\noindent{\sc Proof}: 
There are a few ways to see that there is a short exact sequence
$$0\lrar\bbz\lrar\pi_1(\d{k,k}(C-*))\lrar \bbz^{4g}\lrar 0$$
with generators having the desired properties.\hfill\break
$\bullet$ By considering the map $\mu^2$ in 6.6 
and analyzing the fiber $F$ given by
the complement of a hyperplane in
$\bbc^{k-g}\times\bbc^{k-g}=\bbc^{2(k-g)}$ (the fiber of 
$\mu$ restricted to $\sp{k}(C-*)$
is $\bbc^{k-g}$ as indicated in 6.7). Here $\pi_1(F)=\bbz$.\hfill\break
$\bullet$  Mimicking the proof of Jones for the case $C=\bbp^1$.
One considers the subspace $U$ of $\d{k,k}(C-*))$ consisting of pairs
$(D_0,D_1)$ with the roots of $D_1$ all distinct.  This is the
complement of a (real) codimension $2$ subspace and there is a
surjection $\pi_1(U)\lrar \pi_1(\d{k,k}(C-*))$. Now there is a fibration
$\sp{k}(C-Q_{k+1})\lrar\pi_1(U)\lrar\pi_1(F(C-*, k)$ with
section (where $Q_{k+1}$ is a set of $k+1$ distinct points). 
Since $\pi_1(\sp{k}(C-Q_{k+1}))=H_1(C-Q_{k+1})=\bbz^{2g}\times\bbz^k $,
there is a semi-direct product of $\bbz^{2g}\times\bbz^k$ with the
ordered braid group $B_k(C-*)$ with quotient at least $\bbz^{4g}$.
A similar argument as in [S], lemma 6.4 implies the
answer.
\hfill\za

\noindent{\bf Lemma} 6.11:~{\sl $\pi_1(\bhol{k}(C,\bbp^1))$ has
generators $\tau$ and $e_i,1\leq i\leq g$ which map under 
$\phi_*:
\pi_1(\bhol{k}(C,\bbp^1))\hookrightarrow \pi_1(\d{k,k}(C-*))$ as
follows: $\phi_*(\tau )=\tau$ (the generator of the same name) and
$\phi_*(e_i)=a_ib_i$. In particular $\phi_*$ is an injection. }

\noindent{\sc Proof}: Diagram 6.6 induces a map of fibrations and
since $\pi_2(J)=0$, a map of short exact sequences 
$$\begin{array}{ccccccccc}
  0&\lrar&\bbz&\lrar
  &\pi_1(\bhol{k}(C,\bbp^1))&\lrar&\bbz^{2g}&\lrar 0\\
  &&\decdnar{=}&&\decdnar{\phi_*}&&\decdnar{\Delta_*}&&\\
  0&\lrar&\bbz&\lrar
  &\pi_1(\d{k,k}(C-*))&\lrar&\bbz^{4g}&\lrar 0
\end{array}$$
The map $\Delta_*$ is an injection and then so is the middle map.
The generator $\tau\in\pi_1(\bhol{k})$ 
constructed in 6.4 corresponds exactly to the generator coming from
$\pi_1(F)$ and is described by zeros linking with poles. 
The rest of the claim follows from the commutativity of the above diagram.
\hfill\za

\noindent{\sc Proof of theorem 6.8}: The generators of
$\pi_1$ and their images under $\phi$ having been determined, we next look at
the image of the commutators
$$\phi ([e_i, e_j])=a_i\underbrace{b_ia_j}b_jb_i^{-1}\underbrace{a_i^{-1}
b_j^{-1}}a_j^{-1}$$
Suppose we know that for $k\geq 2$ and $1\leq i, j\leq g$
$$ a_ib_ja_i^{-1}b_j^{-1} = 
  \left\{{\tau,\ \ |i-j|=g}\atop {1,\ \ \hbox{otherwise}}\right.
\in \pi_1(\d{k,k}(C-*)), \leqno{6.12}$$
then $\phi [e_i,e_j ]$ becomes (after transforming the underbraced terms)
and for $|i-j|=g$
\begin{eqnarray*}
    \phi ([e_i, e_j])
    &=&a_i\tau a_jb_ib_j b_i^{-1}b_j^{-1}a_i^{-1}\tau a_j^{-1}\\
    &=&\tau^2a_ja_i [b_i, b_j ]a_i^{-1}a_j^{-1}\\
    &=&\tau^2[a_i, a_j ] = \tau^2
\end{eqnarray*}
This is the claim in 6.8 and therefore the theorem would follow as
soon as we prove 6.12.

Consider the sub-divisor space $\d{k,1}^{2}(C-*))\subset\d{k,k}(C-*))$
consisting of pairs $(D_1,q)\in \sp{k}(C-*)\times (C-*)$ with the 
property that $q\not\in D_1$ (here again $k>1$). 
There is a projection and a fibration
$$\sp{k}(C-\{*, q\})\lrar\d{k,1}(C-*)\lrar (C-*)$$
The generators $b_i\in\pi_1(C-*)$ act on the fundamental group
of the fiber $\pi_1(\sp{k}(C-\{*, q\})$ and this action describes the
commutator map in 6.12. Since $\pi_1(\sp{k}(C-\{*, q\})$ is abelian, we
can look at the action in homology. This has been already determined
in ([C$^2$M$^2$], prop. 10.12) and is exactly given by 6.12 ($b_i$ is
$\tau_{1i}$ in their notation). This completes the proof.
\hfill\za

\noindent{\bf Proposition} 6.13:~{\sl For $k\geq 2g$ and all $n\geq 1$, 
we have an isomorphism 
$$\pi_1(\bhol{k}(C,\bbp^n ))\cong \pi_1(\bmap{0}(C,\bbp^n ))$$}

\noindent{\sc Proof}: When $n>1$, one can show that
$\pi_1(\bhol{k}(C,\bbp^n ))$ is abelian (this has been proven in
[Ki], lemma 3.5). The equivalence in
$\pi_1$ when $n>1$ becomes an equivalence at the level of $H_1$ which
we know is true by Segal's theorem. In fact when $n>1$,
$\pi_1(\bhol{k}(C,\bbp^n ))= \pi_1(\bmap{k}(C,\bbp^n))\cong\bbz^{2g}$
by 6.3(a).

Suppose $n=1$. The proof of 6.11 shows
that the sequence in 6.9 is actually exact.  We would like to see that
6.9 maps to the exact sequence in 6.3(b). The essential point is to show
that the following homotopy commutes
$$\begin{array}{cccc}
  \bhol{k}(C,\bbp^1)&\fract{p}{\ra 7}&\spy (C-*)\\
  \decdnar{i}&&\decdnar{\simeq}\\
  \bmap{k}(C,\bbp^1)&\lrar (\Omega S^2)^{2g}\lrar& 
  (\Omega\bbp^{\infty})^{2g}
\end{array}
$$
where again $p$ sends a holomorphic map to the divisor of its zeros,
the bottom composite $g: \bmap{k}(C,\bbp^1)\lrar
(\Omega\bbp^{\infty})^{2g}$ is restriction to the one skeleton
followed by postcomposition with $S^2\hookrightarrow\bbp^{\infty}$.
Since both spaces on the far right are Eilenberg-Maclane spaces
$(S^1)^{g}$, one simply need consider the effect of $p$ and $g\circ i$
on cohomology (recall that $i$ is an isomorphism on $H^1$ by 1.3).  Label
generators of $H^1((S^1)^{2g})$ by $f_i, 1\leq i\leq 2g$.  As is
explicit in theorem 4.13, the $g^*(f_i)$ correspond to the non-trivial
torsion free classes in $H^1(\bmap{k})$ (see 4.13 and 4.9).  Similarly
the construction of the stable classes in \S5 (see example 5.4)
implies precisely that $p^*(f_i)$'s are the torsion free generators of
$H^1(\bhol{k} )$.  That is $p^*=i^*\circ g^*$ and hence $p\simeq
i\circ g$.

Now and upon applying $\pi_1$ to the terms of the diagram above as
well as to those in 6.5 we get a map from 6.9 down to 6.3(b). Since
the end terms of the sequences are isomorphic, it follows by the five
lemma that the middle terms are isomorphic; i.e.
$\pi_1(\bhol{k})\cong\pi_1(\bmap{k})$.  \hfill\za

\noindent{\bf Corollary} 6.14:~{\sl $\pi_1(\bmap{0}(C,\bbp^n ))$ is
generated by classes $e_1,\ldots, e_{2g}$ and $\alpha$ such that the
commutators $[e_i, e_{g+i}] = \alpha^2$ and all other commutators are
zero. }

\noindent{\sc Note}: Proposition 6.13 is not sufficient to imply that
Segal's homology equivalence 5.7 can be upgraded to a homotopy
equivalence. A detailed study of the actions of $\pi_1$ on the
universal covers is needed.

\vskip 10pt
\noindent{\bf\Large\S7 Spaces of Unbased Maps}
\vskip 10pt

There is an important distinction between $\map{}^*(C, X)$ and
$\map{}(C, X)$. For one thing, while the topology of the based mapping
space doesn't vary with the degree, this is no longer true in the
unbased case. Let $\map{f}(C_g,X)$ denote the component containing
a given map $f$,
$x_0\in C_g$ the basepoint and consider again the ``evaluation" fibration
$$\bmap{f}(C_g,X)\lrar\map{f}(C_g,X)\fract{ev}{\lrar} X~,~~ev(g) =
g(x_0)\leqno{7.1}$$
Associated to 7.1 is a long exact sequence of homotopy groups
$$\rightarrow\pi_i\bmap{f}(C_g,X)\rightarrow
\pi_i\map{f}(C_g,X)\rightarrow\pi_i (X)
\fract{\partial}{\lrar}\pi_{i-1}\bmap{f}(C_g,X)\rightarrow$$
Suppose $X$ is simply connected. Notice that given a map $f:C_g\lrar X$,
$f$ is null on $\bigvee S^1$ and hence factors (up to homotopy) as in
$$f: C_g\lrar S^2\fract{{\bar f}}{\ra 3}X$$
One then has the following diagram of vertical fibrations over $X$
$$\begin{array}{ccccc}
  \Omega_{\bar f}^2X&\lrar&\bmap{f}(C_g,X)&\lrar& (\Omega X)^{2g}\\
  \decdnar{}&&\decdnar{}&&\decdnar{}\\
  {\cal L}^2_{\bar f}X&\lrar&\map{f}(C_g,X)&\lrar& \map{}(\bigvee S^1, X)\\
  \decdnar{ev}&&\decdnar{ev}&&\decdnar{ev}\\
  X&\fract{=}{\lrar}&X&\fract{=}{\lrar}&X
\end{array}
\leqno{7.3}$$
inducing a diagram of homotopy groups and boundary homomorphisms
$$\begin{array}{ccccc}
  \pi_i(X)&\fract{=}{\lrar}&\pi_i(X)&\fract{=}{\lrar}&\pi_i(X)\\
  \decdnar{\partial_1}&&\decdnar{\partial}&&\decdnar{\partial_2}\\
  {{\pi_{i-1}(\Omega^2(X))}\atop{=\pi_{i+1}(X)}}
  &\lrar&\pi_{i-1}(\bmap{f}(C_g,X))&\lrar&
  {{\pi_{i-1}(\Omega_{fi} X)^{2g})}\atop {= \pi_i(X)^{2g}}}
\end{array}
$$

\noindent{\bf Proposition} 7.4:~{\sl Assume $X$ simply connected,
$f: C_g\lrar X$ and $\bar f$ such that
$C_g\fract{\pi}{\lrar}S^2\fract{\bar f}{\lrar}X$ (up to homotopy).
Assume $i>2$. Then the homotopy boundary
$$\partial:\pi_i(X)\lrar\pi_{i-1}(\bmap{f}(C_g,X))=\pi_{i}(X)^{2g}\oplus
\pi_{i+1}(X)$$
for the fibration
$\bmap{f}(C_g,X)\lrar\map{f}(C_g,X)\fract{ev}{\lrar} X$
factors through the term $\pi_{i+1}(X)$ and
is given (up to sign) by the Whitehead product:
$\partial \alpha = [\alpha,{\bar f}].$}

\noindent{\sc Proof:}
The sequence
$0\rightarrow\pi_{i+1}(X)\rightarrow\pi_{i-1}(\bmap{f}(C_g,X))
\rightarrow\pi_i(X)^{2g}\rightarrow 0$ splits canonically (according to 6.2)
and with respect to that splitting one has
$\partial=\partial_1+\partial_2$. By Whitehead's theorem we have that
$\partial_1\alpha = [\alpha, {\bar f}]$ while $\partial_2\alpha = 0$.
The proof follows.
\hfill\za

\noindent{\sc Remark}: When $i=2$ the situation has to be handled
differently for $\pi_{1}(\bmap{f}(C_g,X))$ does not necessarily split
as in 7.4.

\noindent{\bf Lemma} 7.5: 
$\pi_1(\map{d}(C,\bbp^n ))\cong\bbz^{2g}$
when $n>1$.

\noindent{\sc Proof:}
   Consider the Hopf fibration $S^{2n+1}\lrar\bbp^n\lrar
\bbp^{\infty}$ and the induced fibration
$$\map{}(C, S^{2n+1})\lrar\map{d}(C,\bbp^n
)\lrar\map{d}(C,\bbp^{\infty})
\simeq(S^1)^{2g}\times\bbp^{\infty}$$
Since $\pi_1(\map{}(C, S^{2n+1}))=0$ when $n>1$, the result follows. 
\hfill\za

We next address the case $n=1$ and analyze 7.3 when
$X=\bbp^1$. Let $f:S^2\lrar X$ be a degree $d$ map.
The boundary term in the left vertical fibration
$\pi_1(\Omega^2_fX )\fract{\partial}{\lrar}\pi_1({\cal L}^2_f(X))$
is given according to 7.4 by multiplication by $2d$. On the other
hand, the right vertical fibration has a section and hence we get the
diagram of fundamental groups
$$\begin{array}{ccccccccc}
 0&\lrar&\bbz&\lrar&\pi_1(\map{d}^*)&\lrar&\bbz^{2g}&\lrar& 0\\
 \decdnar{}&&\decdnar{}&&\decdnar{}&&\decdnar{=}&&\decdnar{}\\
 0&\lrar&\bbz_{2d}&\lrar&\pi_1(\map{d})&\lrar&\bbz^{2g}&\lrar&0
\end{array}
$$

\noindent{\bf Corollary} 7.6:~(Larmore-Thomas) {\sl $\pi_1(\map{d}(C,
S^2 ))$ is generated by classes $e_1,\ldots, e_{2g}$ and $\alpha$ such
that
$$\alpha^{2|d|}=1,~~ [e_i, e_{g+i}] = \alpha^2$$
All other commutators are zero. In particular $\map{d}(C,\bbp^1 )$ and
$\map{d'}(C, \bbp^1)$ have different homotopy types whenever $d\neq
\pm d'$.}

\noindent{\bf Corollary} 7.7:~{\sl For $d\geq 2g$, $\pi_1(\hol{d}(C,
\bbp^1 ))$ is generated by $e_1,\ldots, e_{2g}$ and $\alpha$
with $\alpha^{2|d|}=1,~~ [e_i, e_{g+i}] = \alpha^2$ and all other
commutators are zero.}

\noindent{\sc Proof}: This is equivalent to showing that
$\pi_1(\hol{d}(C, \bbp^1 ))$ corresponds to $\pi_1(\map{d}(C,
S^2 ))$ in that range but this follows from a direct comparison of
the evaluation fibrations for both mapping spaces and theorem 1.4.
\hfill\za

In the case of maps into $\bbp^n$, $n\geq 2$, the fundamental group
is not enough to distinguish between connected components. Proposition
7.4 still implies

\noindent{\bf Proposition} 7.8:~{
$\map{k}(C_g,\bbp^{2d})$ and
$\map{l}(C_g,\bbp^{2d})$ have different homotopy types
whenever $l\not\equiv k (2)$.}

\noindent{\sc Proof:} 
The long exact sequence for the evaluation fibration gives again
$$\cdots\pi_{2d+1}\bbp^{2d}\fract{\partial}{\lrar}
\pi_{2d}(\bmap{k})\rightarrow
\pi_{2d}(\map{k} )\rightarrow\pi_{2d}(\bbp^{2d})\fract{\partial}{\lrar}
\cdots$$
where $\bmap{k}$ stands for $\bmap{}(C_g,\bbp^{2d})$.
According to 7.4 the sequence above becomes
$$\lrar \pi_{2d+1}\bbp^{2d}=\bbz\langle\eta\rangle
\fract{k[\iota, \eta]}{\ra 4}\pi_{2d+2}(\bbp^{2d})=\bbz_2\rightarrow
\pi_{2d}(\map{k} )\rightarrow 0$$
where $\iota$ is the generator of $\pi_2(\bbp^{2d})$.
The Whitehead product pairing for complex
projective spaces
$$\pi_{2m+1}(\bbp^m)\tensor\pi_2(\bbp^m)\lrar\pi_{2m+2}(\bbp^m)$$
is worked out in [P]. The result there is that the Whitehead product
is zero if $m$ is odd and non-zero if $m$ is even. Since we are in the
case $m=2d$, the map 
$\bbz\langle\eta\rangle\fract{k[\iota, \eta]}{\ra 3}\bbz_2$ is in fact
multiplication by $k$ and the proposition follows right away.
\hfill\za


\vskip 10pt

\addcontentsline{toc}{section}{Bibliography}
\bibliography{biblio}

\begin{thebibliography}{doC}
\bibitem{}[K4] S. Kallel, ``An analog of the May-Milgram model for
configurations with multiplicities'', this volume.
\end{thebibliography}
\bibliographystyle{plain}


\end{document}